\documentclass[showpacs,preprintnumbers,amsmath,amssymb]{revtex4}
\usepackage{graphicx}
\usepackage{dcolumn}
\usepackage{bm}

\begin{document}
\draft

\title{Engineering Electromagnetic Properties of Periodic Nanostructures Using
Electrostatic Resonances}

\author{Gennady Shvets and Yaroslav Urzhumov}
\affiliation{Department of Physics, The University of Texas at
Austin, Austin, Texas 78712}

\newcommand{\ba}{\begin{eqnarray}}
\newcommand{\ea}{\end{eqnarray}}
\newcommand{\be}{\begin{equation}}
\newcommand{\ee}{\end{equation}}
\newcommand{\para}{\parallel}

\begin{abstract}
Electromagnetic properties of periodic two-dimensional
sub-wavelength structures consisting of closely-packed inclusions
of materials with negative dielectric permittivity $\epsilon$ in a
dielectric host with positive $\epsilon_h$ can be engineered using
the concept of multiple electrostatic resonances. Fully
electromagnetic solutions of Maxwell's equations reveal multiple
wave propagation bands, with the wavelengths much longer than the
nanostructure period. It is shown that some of these bands are
described using the quasi-static theory of the effective
dielectric permittivity $\epsilon_{qs}$, and are independent of
the nanostructure period. Those bands exhibit multiple cutoffs and
resonances which are found to be related to each other through a
duality condition. An additional propagation band characterized by
a negative magnetic permeability develops when a magnetic moment
is induced in a given nano-particle by its neighbors. Imaging with
sub-wavelength resolution in that band is demonstrated.
\end{abstract}


\maketitle

Electrostatic resonances of isolated nanoparticles have recently
attracted substantial interest because of the intriguing
possibility of obtaining very strong and localized electric
fields. Applications of such fields include nanoscale
biosensors~\cite{vanduyne_jacs02,akhremitchev_lang02},
nanolithography~\cite{alkaisi_apl99,vanduyne_jpc99}, and nonlinear
spectroscopy~\cite{moskovits_chem00,halas_apl03}. Resonances occur
when a metallic or dielectric particle with a negative
frequency-dependent dielectric permeability $\epsilon(\omega) < 0$
is imbedded in a dielectric host (including vacuum) with a
positive dielectric permeability $\epsilon_h > 0$. The wavelengths
$\lambda$ of the incident electromagnetic radiation that resonate
with a small particle of a characteristic size $d \ll \lambda$
depend on the particle shape and the functional dependence of
$\epsilon(\omega)$. By changing the shape~\cite{fredkin03} and
internal composition~\cite{halas_apl03} of nanoparticles
resonances can be shifted to the wavelength optimized for a
particular application. Close proximity of other small particles
can also strongly affect the resonances. For example, disordered
conglomerates of nanoparticles have recently been analyzed using
the quasi-static theory~\cite{bergman_stroud92} and
shown~\cite{stockman_bergman01} to exhibit multiple coherent "hot
spots" that can be used for transporting electromagnetic energy.
Surface-enhanced Raman scattering (SERS) from molecules inside
multi-particle aggregates has also been shown to be greatly
enhanced~\cite{brus_jpc00,markel_prb99,shalaev_nano04}.

In this Letter we demonstrate how electrostatic resonances of {\it
periodic} arrays of closely-spaced particles can be used for
engineering propagation properties of electromagnetic radiation
through such nanophotonic crystals. The emphasis of this work is
on subwavelength photonic crystals (SPC) that have a periodicity
$d$ significantly smaller than the radiation wavelength in vacuum
$\lambda \equiv 2\pi c/\omega$. By numerically solving Maxwell's
equations, we identify two classes of waves supported by an SPC:
(a) hybridized Dipole Modes (DM) that are characterized by a
quasi-static period-independent dielectric permittivity
$\epsilon_{qs}(\omega)$, and (b) hybridized Higher-order Multipole
Modes (HMM) that depend on the crystal period $d$. Two types of
DM's are identified: almost dispersionless (non-propagating)
collective plasmons (CPL) satisfying the $\omega(\vec{k}) \equiv
\omega_i^{(c)}$ dispersion relation (where $\omega_i^{(c)}$ are
multiple zeros of $\epsilon_{qs}$), and propagating collective
photons (CPH) satisfying the $\vec{k}^2c^2 = \omega^2
\epsilon_{qs}(\omega)$ dispersion relation. The mean-field
dielectric permittivity $\epsilon_{qs}$ calculated from the
quasi-static theory~\cite{bergman_stroud92,stockman_bergman01} is
found to be highly accurate in predicting wave propagation even
for SPCs with the period as large as $\lambda/2\pi$. DM wave
propagation bands are "sandwiched" between multiple resonance
$\omega_i^{(r)}$ and the cutoff $\omega_i^{(c)}$ frequencies of
the SPC. For two-dimensional SPC's with a high lattice symmetry
(square and hexagonal) a duality condition expressing a simple
one-to-one correspondence between the resonant and cutoff
frequencies is derived.

The new HMM propagation bands are discovered inside the frequency
intervals where $\epsilon_{qs} < 0$ and, by the mean-field
description, propagation is prohibited. HMM bands should not be
confused with the usual high order Brillouin zones of a photonic
crystal because the latter do not satisfy the $d \ll \lambda$
condition. One HMM band defines the frequency range for which the
sub-wavelength photonic crystal behaves as a double-negative
metamaterial (DNM) that can be described by the negative effective
permittivity $\epsilon_{eff} < 0$ and permeability $\mu_{eff} <
0$~\cite{smithschultz_prl00}. Magnetic properties of the DNM are
shown to result from the induced magnetic moment inside each
nanoparticle by high-order multipole elecctrostatic resonances of
its neighbors. It is shown that a thin slab of such DNM can be
employed as sub-wavelength lens capable of resolving images of two
slits separated by a distance $\ll \lambda$.

To start, consider a TM-polarized electromagnetic wave, with
non-vanishing $H_z$, $E_x$, and $E_y$ components, incident on an
{\it isolated} dielectric rod (infinitely long in the
$z-$direction) with $\epsilon(\omega) < 0$. The incident em wave
is strongly scattered by the rod when its frequency $\omega$
coincides with that of the surface plasmon found by solving the
nonlinear eigenvalue equation for $H_z$:
\begin{equation}\label{eq:hz_master}
    -\vec{\nabla} \cdot \left( \epsilon^{-1} \vec{\nabla} H_z
    \right) = \frac{\omega^2}{c^2} H_z,
\end{equation}
where $H_z \rightarrow 0$ far from the rod. The electric field of
the surface wave is given by $\vec{E}(\vec{x}) = -i [c/\omega
\epsilon(\vec{x})] \vec{e}_z \times \vec{\nabla} H_z$, where
$\epsilon(\vec{x}) = \epsilon_h$ outside and $\epsilon(\vec{x}) =
\epsilon(\omega)$ inside the rod. In what follows we assume that
the rods are in vacuum, i.~e.~$\epsilon_h = 1$. Note that the
lines $H_z = {\rm const}$ are the electric field lines. For a
sub-wavelength rod the rhs of Eq.~(\ref{eq:hz_master}) can be
neglected. Moreover, $|\vec{E}| \gg |H|$, and the description
using the electrostatic potential $\phi$ is appropriate: $\vec{E}
= -\vec{\nabla} \phi$. Hence, two equations are simultaneously
satisfied for a sub-wavelength rod:
\begin{equation}\label{eq:duality}
    -\vec{\nabla} \cdot \left( \epsilon^{-1} \vec{\nabla} H_z \right)
    = 0 \ \ \ {\rm and} \ \ \  -\vec{\nabla} \cdot
    \left( \epsilon \vec{\nabla} \phi \right) = 0.
\end{equation}
Equations (\ref{eq:duality}) illustrate that in electrostatics
there are two equivalent descriptions of the electric field:
potential description and field line description. Note from
Eqs.~(\ref{eq:duality}) that if a surface wave is supported by a
rod of an arbitrary transverse shape for $\epsilon = \epsilon_1$,
then a surface wave is also supported for $\epsilon =
1/\epsilon_1$~\cite{fredkin03}.

\begin{figure}
\includegraphics[height=3.2cm]{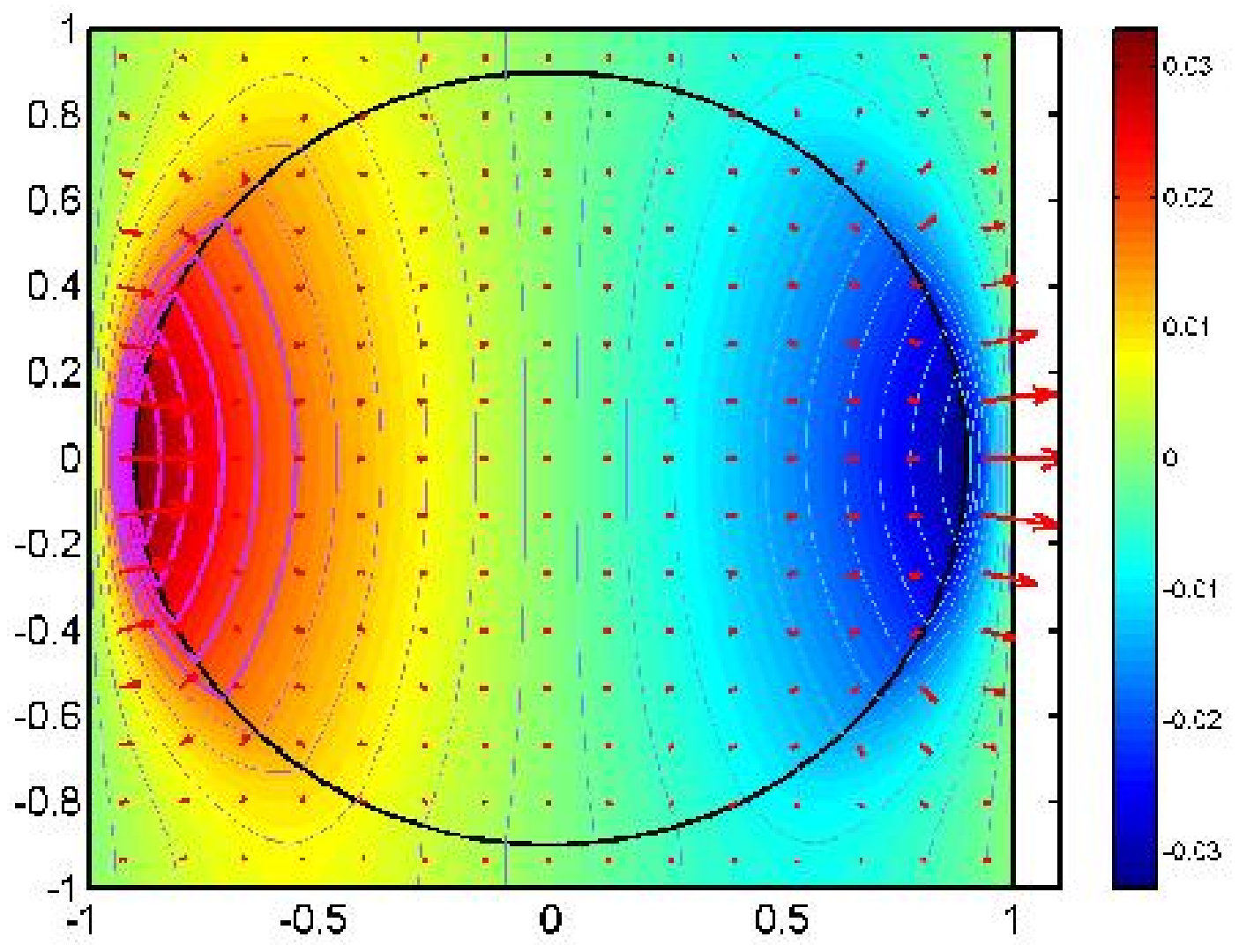}
\includegraphics[height=3.2cm]{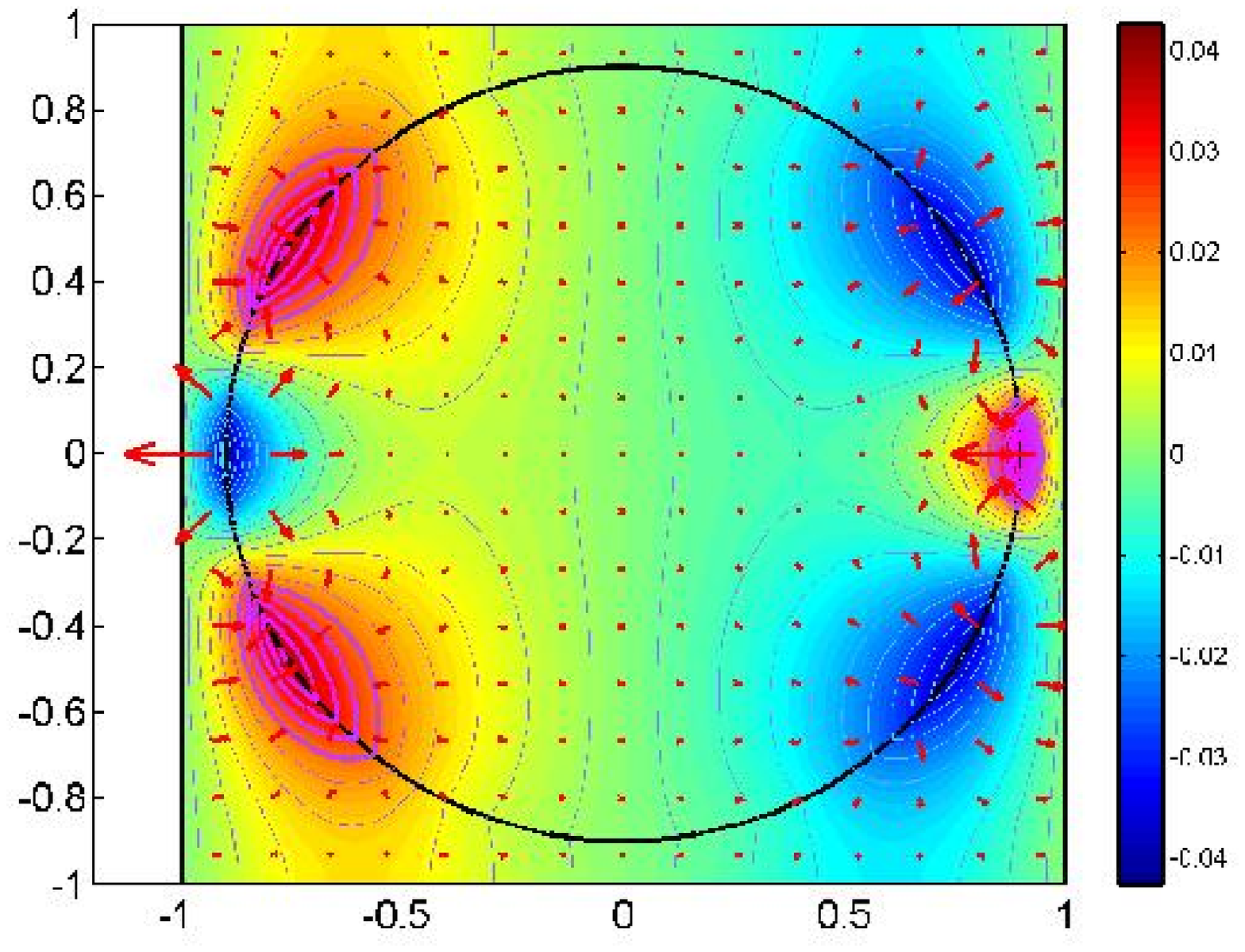}
\caption{\label{fig:two_resonances} Two strongest hybridized
electrostatic dipole  resonances of the 2-D square lattice of
plasmonic rods contributing to $\epsilon_{qs}$: $\omega_1^{(r)} =
0.38 \omega_p$ (left); $\omega_2^{(r)} = 0.63 \omega_p$ (right).
Shown are the potential isocontours and the electric field arrows.
Spatial dimensions are measured in $2c/\omega_p$ units. SPC
parameters: $d = c/\omega_p$ and $R = 0.45 d$.}
\end{figure}

\begin{figure}
\includegraphics[height=5.2cm]{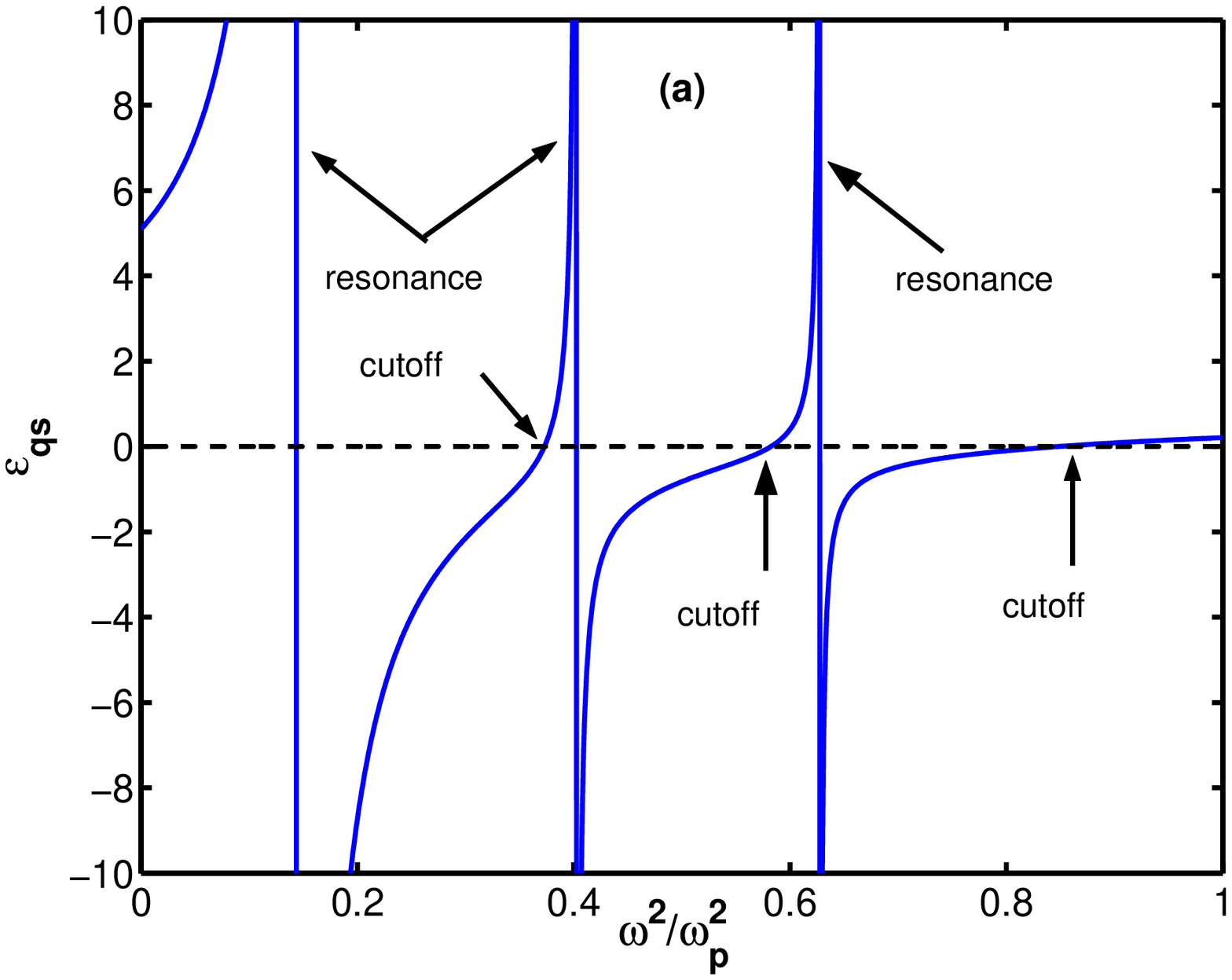}
\includegraphics[height=5.2cm]{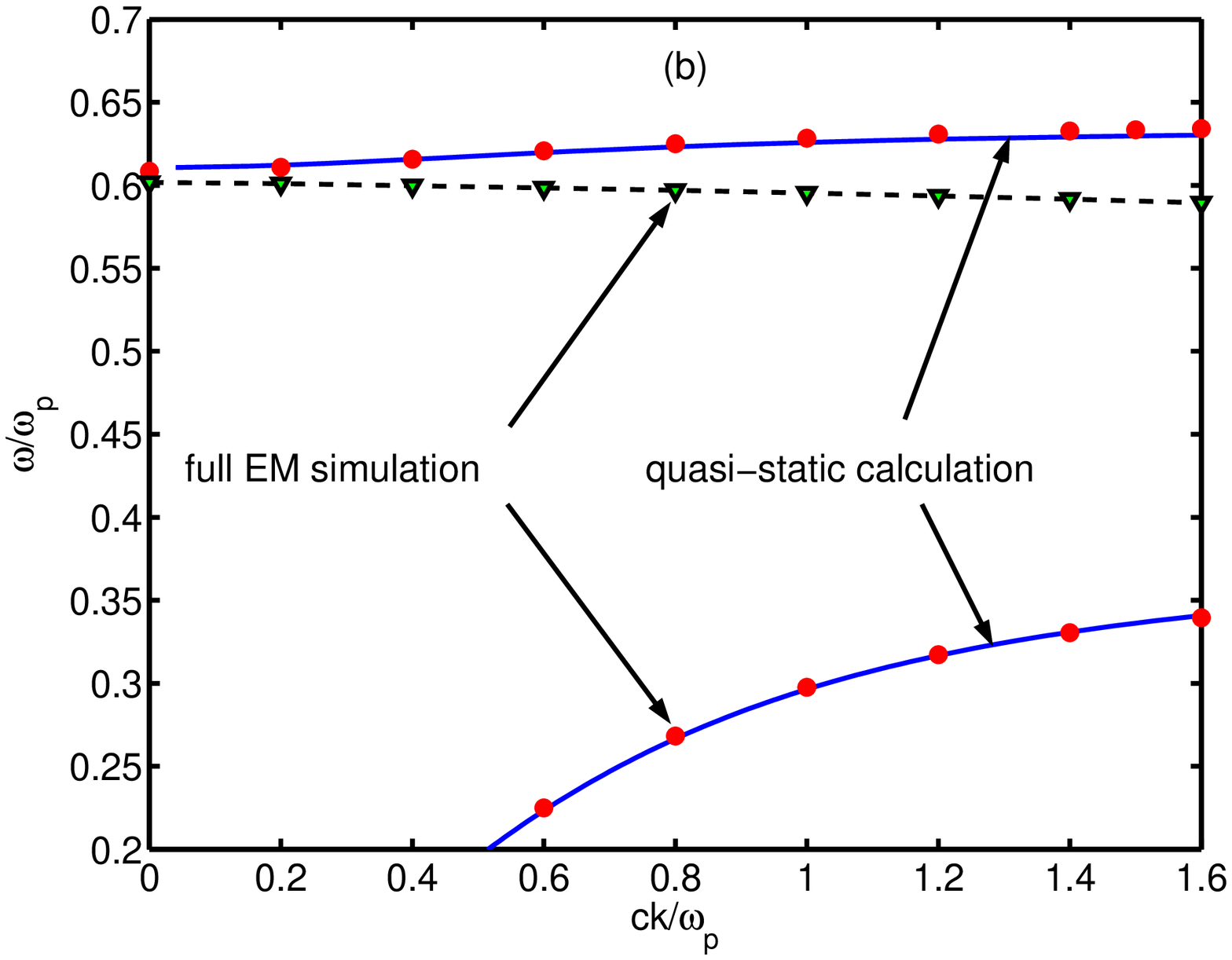}
\caption{\label{fig:epseff} (a) Frequency dependence of the
quasi-static dielectric permittivity $\epsilon_{qs}$ from
Eq.~(\ref{eq:eps_eff1}). Vertical lines: resonances, horizontal
axis intercepts: cutoffs. (b) Collective photons (CPH) supported
by the SPC: theoretical prediction of the quasi-static theory, $k
= \sqrt{\epsilon_{qs}} \omega/c$ (solid lines) and numerical
simulation of Eq.~(\ref{eq:hz_master}) (circles and diamond). Note
the backward-wave mode ($\partial \omega/\partial k < 0$) marked
by diamonds which is not described by the quasi-static theory. SPC
parameters as in Fig.~\ref{fig:two_resonances}.}
\end{figure}

For a round cylinder of radius R the surface-wave multipole
solutions of the second of Eqs.~(\ref{eq:duality}) are given by
$\phi_{out}^{(m)} = (r/R)^{-m} \exp{(im\theta)}$ outside and
$\phi_{in}^{(m)} = (r/R)^{m} \exp{(im\theta)}$ inside the
cylinder, and $m \geq 1$. The continuity of $\epsilon \partial
\phi/\partial r$ is satisfied for every $m \geq 1$ for
$\omega^{(r)}$ such that $\epsilon(\omega^{(r)}) = -1$. This
degeneracy of the multipole resonances is specific to the 2-D
lattice of round cylinders. This property makes the square SPC of
round cylinders particularly amenable to engineering its
electromagnetic properties using electrostatic resonances: various
multipoles of a given rod are strongly hybridized by the proximity
of other rods. This gives rise to a rich and easily controllable
set of hybridized electrostatic resonances whose frequencies
depend on the areal fraction occupied by the cylinders.
Earlier approximate
calculations~\cite{lamb_wood80} of the three-dimensional lattice
of metallic spheres failed to demonstrate the type of rich
multi-resonant band structure found for 2-D structures and shown
in Fig.~\ref{fig:epseff} for two reasons: (a) there is no
frequency degeneracy in three dimensions, and (b) the volume
fraction occupied by metallic spheres is 3-D is smaller than in
2-D.

Note that there is no monopole ($m=0$) {\it electrostatic}
resonance (although there is an electromagnetic Mie resonance for
rods with a very high positive
$\epsilon$~\cite{pendryobrien_jop02}) for an isolated cylinder.
Such a resonance would reveal that the corresponding solution for
$H_z$ has an associated azimuthal $\theta$-independent current
around the cylinder and, thus, a non-vanishing magnetic moment $M
= (1/2c) \langle \vec{r} \times \vec{J} \rangle$, where the
average is taken over the unit cell. However, as shown below, the
octupole ($m = 4$) electrostatic resonances in a square lattice of
closely-packed cylinders hybridize in a way of inducing a
resonantly excited magnetic moment. This magnetic moment manifests
itself as a negative effective permeability $\mu_{eff}$ of the
structure, and an additional propagation band of DNW's in a narrow
frequency range.

In the rest of the paper we concentrate on a specific SPC: a
square lattice of round ($R = 0.45 d$) {\it plasmonic} cylinders
with $\epsilon(\omega) = 1 - \omega_p^2/\omega^2$ characteristic
of collisions-free electron gas, lattice period $d = c/\omega_p$.
Very similar results are expected for polaritonic rods with
$\epsilon(\omega) = \epsilon_{\infty} (\omega^2 -
\omega_{LO}^2)/(\omega^2 - \omega_{TO}^2)$, with $\epsilon < 0$
for $\omega_{TO} < \omega < \omega_{LO}$. To this SPC we apply the
standard procedure~\cite{bergman_stroud92,stockman_bergman01} for
calculating the quasi-static dielectric permittivity
$\epsilon_{qs}(\omega)$, and later compare the band structure
described by $\epsilon_{qs}(\omega)$ to that obtained by solving
the fully-electromagnetic Eq.~\ref{eq:hz_master}.

The material-independent $\epsilon_{qs}$ is
calculated~\cite{bergman_stroud92,stockman_bergman01} as
\begin{equation}\label{eq:eps_eff1}
    \epsilon_{qs} = 1 - p \sum_{i=1}^{N} \frac{F_i}{s - s_i},
\end{equation}
where, for the plasmonic rods in vacuum, $s(\omega) = [1 -
\epsilon(\omega)]^{-1} = \omega^2/\omega_p^2$, $s_i \equiv [1 -
\epsilon_i]^{-1}$ is the $i$'th out of $N>1$ hybridized dipole
resonances, $F_i$ its oscillator strength calculated below, and $p
= \pi R^2/d^2$. Electrostatic resonances $s_i$ are found by
solving the eigenvalue equation for the potential eigenfunctions
$\phi_i$ inside a unit cell of the structure:
\begin{equation}\label{eq:eigen_phi1}
    \vec{\nabla} \cdot \left[ \theta(\vec{x}) \vec{\nabla} \phi_i
    \right] = s_i \nabla^2 \phi_i,
\end{equation}
where $\phi_i$ satisfies the following homogeneous conditions at
the unit boundaries $(x,y) = (\pm d/2, \pm d/2)$: (a) $\phi_i$ and
its derivatives are periodic; (b) $\phi_i(x=\pm d/2) = 0$; (c)
$\partial_y \phi_i(y=\pm d/2) = 0$. Equation (\ref{eq:eigen_phi1})
follows from Eq.~(\ref{eq:duality}). Physically, these
eigenfunctions describe the electric potential distribution when a
vanishing ac voltage (with frequency $\omega$ such that
$\epsilon(\omega) = \epsilon_i$) is applied between $x=\pm d/2$
capacitor plates.
The capacitance of such an imaginary capacitor,
equal to the ratio of the charge to the voltage drop, is given by
$C = \epsilon_{qs} d$, and becomes infinite according to
Eq.~(\ref{eq:eps_eff1}).
Another eigenfunction $\tilde{\phi}_i$ corresponding to the
voltage drop between $y=\pm d/2$ plates is obtained by a
$90-$degree spatial rotation of $\phi_i$.

Because the square lattice is invariant with respect to the
transformations of the $C_{4v}$ point group~\cite{lyubarskii}, all
periodic solutions transform according to one of the irreducible
representations (irreps) of $C_{4v}$: four singlets (commonly
labeled as $A_1$, $A_2$, $B_1$, and $B_2$) and one doublet $E$.
The electrostatic eigenfunctions $\phi_i$ and $\tilde{\phi}_i$
have the symmetry of $E$. Inside a given rod each $\phi_i$ can be
expanded as the sum of multipoles: $\phi_i(r,\theta) =
\sum_{l=0}^{\infty} A_i^{(2l+1)} (r/R)^{2l+1} \cos{(2l+1)\theta}$.
A straightforward calculation following
Ref.~\cite{stockman_bergman01} yields the oscillator strength
proportional to the dipole component of $\phi_i$: $F_i = \left(
A_i^{1} \right)^2 /\sum_{l=0}^{\infty} (2l + 1) \left( A_i^{2l+1}
\right)^2$. For our structure there are three significantly strong
hybridized dipole ($E-$symmetric) resonances : $(s_1 = 0.1433, F_1
= 0.8909)$, $(s_2 = 0.4025, F_2 = 0.064)$, and $(s_3 = 0.6275, F_3
= 0.0366)$.

The plots of the two lowest resonances are shown in
Fig.~\ref{fig:two_resonances}. The first resonance is primarily
dipolar ($\propto \cos{\phi}$) while the second one has a
significant sextupolar ($\propto \cos{3 \phi}$) component. Thus,
the close proximity of the rods in the lattice results in a strong
hybridization of the odd multipoles with the dipole. Moreover, the
hybridized dipole resonances $\omega_1^{(r)} = 0.38 \omega_p$,
$\omega_2^{(r)} = 0.63 \omega_p$, and $\omega_3^{(r)} = 0.79
\omega_p$ occur at the frequencies {\it controllably} different
(through the $R/d$ ratio) from that of an isolated rod,
$\omega^{(r)} = \omega_p/\sqrt{2}$.

The frequency dependence of the $\epsilon_{qs}$ for a plasmonic
material is plotted in Fig.~\ref{fig:epseff}(a).
The vertical
lines in Fig.~\ref{fig:epseff}(left) are the plasmonic structure
resonances which occur at $\omega_{1}^{(r)} = 0.38 \omega_p$,
$\omega_{2}^{(r)} = 0.63 \omega_p$, and $\omega_{3}^{(r)} = 0.79
\omega_p$.
Quasi-static theory predicts no propagation in the region of
$\epsilon_{qs} < 0$.
Three frequencies $\omega_{i}^{(c)}$ for which
$\epsilon_{qs}(\omega_{i}^{(c)}) = 0$ are called the cutoff
frequencies. From Fig.~\ref{fig:epseff}(a), there are four
propagation bands (where $\epsilon_{qs} > 0$) allowed by the
quasi-static theory for $\omega < \omega_p$. The quasi-static
theory predicts that this plasmonic SPC acts as an effective
medium supporting two types of waves: four collective photons
(CPHs) satisfying the $|\vec{k}|^2 = \epsilon_{qs} \omega^2/c^2$
dispersion  relation (two of which are indicated by a solid line
in Fig.~\ref{fig:epseff}(b)), and three non-propagating collective
plasmons (CPLs) satisfying the $\omega(\vec{k}) =
\omega_{i}^{(c)}$ dispersion relation (not shown). Numerical
solutions of the fully electromagnetic Eq.~(\ref{eq:hz_master})
described below confirm these conclusions, yet reveal additional
modes whose behavior {\it is not} described by the quasi-static
$\epsilon_{qs}$.

An important duality principle for the cutoff and resonance
frequencies exists for square and hexagonal lattices
(correspondingly invariant with respect to the $C_{4v}$ and
$C_{6v}$ point group transformations): for each resonance
frequency $\omega_{j}^{(r)}$ there exists a cutoff frequency
$\omega_{i}^{(c)}$ for which $s\left( \omega_{i}^{(c)} \right) =
1$. This principle follows from an observation that, at the
cutoff, $H_z$ satisfying the first Eq.~(\ref{eq:duality}) obeys
the same boundary conditions as $\phi$ and $\tilde{\phi}$. Hence,
from Eq.~(\ref{eq:duality}) follows that for each resonance
frequency $\omega_{j}^{(r)}$ there exists a cutoff frequency
$\omega_{i}^{(c)}$ for which $\epsilon(\omega_{i}^{(c)}) =
1/\epsilon(\omega_{j}^{(r)})$. Numerical calculation of the zeros
of $\epsilon(s)$ from Eq.~(\ref{eq:eps_eff1}) indeed confirms the
duality principle.

To verify that $\epsilon_{qs}$ is sufficient for accurate
description of em wave propagation through the plasmonic SPC, the
fully electromagnetic Eq.~(\ref{eq:hz_master}) was numerically
solved as a nonlinear eigenvalue equation for
$\omega^2/\omega_p^2$ for different wavenumbers $\vec{k} =
\vec{e}_x k$, and the resulting dispersion relation $\omega(k)$
plotted in Fig.~\ref{fig:epseff}(b) for three propagation bands.
Although $d = 1/\omega_p$ is not infinitesimally small compared to
the radiation wavelength, it is apparent from
Fig.~\ref{fig:epseff} that the numerically calculated points
(circles) accurately fall on the solid lines predicted for the
CPH's by the scale-independent $\epsilon_{qs}$. The essentially
flat propagation band $\omega(k) \approx 0.61 \omega_p$ not shown
in Fig.~\ref{fig:epseff}(b) also agrees with the CPL dispersion
relation obtained from $\epsilon_{qs}$. Therefore, for several
frequency bands, the plasmonic SPC indeed is an effective medium
described by the scale-independent $\epsilon_{qs}$.

Importantly, Fig.~\ref{fig:epseff}(b) reveals that there is
another propagation band (diamonds) in the frequency range for
which no propagation is expected due to $\epsilon_{qs} < 0$. Note
that the mode's group velocity $v_g=\partial \omega/\partial k< 0$
opposes its phase velocity -- an indication that we're dealing
with a DNM. For $\vec{k} = 0$ this mode's $H_z$ has the symmetry
of the $A_1$ irrep of the symmetry group $C_{4v}$, and can be
expanded inside a given plasmonic rod as $H_z(r,\theta) =
\sum_{k=0}^{\infty} A^{(4k)} \left[ I_{4k}(\omega \sqrt{-\epsilon}
r/c)/I_{4k}(\omega \sqrt{-\epsilon} R/c) \right]
\cos{(4k\theta)}$, where $I_{l}$ is the modified Bessel function
of order $l$.
Because there is no dipole component in $H_z$, the
$A_1$ mode does not contribute to the quasi-static permittivity
$\epsilon_{qs}$.
For the SPC at hand, the largest term in the expansion is the
octupole term $A^{(4)}$, and the next largest is the monopole term
$A^{(0)}$ that is responsible for the magnetic moment induced in
the photonic structure as explained earlier. Therefore, the mode
is an HMM, with predominantly $m = 4$ component. Because of the
finite magnetic moment, a single quantity $\epsilon_{qs}$ cannot
describe the HMR resonances, and two frequency-dependent
parameters are numerically evaluated: dielectric permittivity
$\epsilon_{eff}$ and magnetic permeability $\mu_{eff}$. The
procedure for expressing these effective quantities for a periodic
structure using the cell-averaged electric and magnetic fields has
been described elsewhere~\cite{pendry_ring99,pendryobrien_jop02}.

For two dimensions, and assuming that the elementary cell of the
SPC is centered at the origin, we introduce several variables by
averaging $H_z$ and $\vec{E}$ over the sides or the area of the
elementary cell of the photonic crystal: $\tilde{B}_z =d^{-2} \int
dA H_z(x,y)$, $\tilde{H}_z = H_z(x=-d/2,y=-d/2)$,
$\tilde{E_y}=d^{-1} \int_{-d/2}^{+d/2} dy \ E_y(x=-d/2,y)$, and
$\tilde{D}_y = d^{-1} \int_{-d/2}^{d/2} dx E_y(x,y=-d/2)$. For
$\vec{k}=k\vec{e}_x$ Maxwell's equations in the integral form
become $k c g(kd) \tilde{E}_y = \omega \tilde{B}_z$ and $kc g(kd)
\tilde{H}_z = \omega \vec{D}_y$, in complete correspondence with
Maxwell's equations in the medium for which $\tilde{B}_z =
\mu_{eff} \tilde{H}_z$ and $\tilde{D}_y = \epsilon_{eff}
\tilde{E}_y$. Dimensionless factor $g(x) = i [1 - \exp{(ix)} ]/x
\rightarrow 1$ for $kd \ll 1$ is the slight modification
accounting for non-vanishing lattice period.

The magnetic permeability $\mu_{eff}$ is affected because the mode
carries the electric current which produces a finite magnetic
moment. The magnetic nature of the $A_1$ mode is due to the
non-vanishing coefficient of the monopole $A^{(0)}$ term in the
multipole expansion. The monopole is responsible for the
$\theta$-independent component of the azimuthal electric field
$E_{\theta} = -i[c/\omega \epsilon(r)]\partial_r H_z$. The
corresponding electric current in the negative-$\epsilon$ rod
given by $J_{\theta} = -A^{(0)} \omega/c (1 - 1/\epsilon)
I_1(\omega r/c)/I_0(\omega r/c)$ produces a magnetic moment
density $\vec{M} =  (1/2c) \langle \vec{r} \times \vec{e}_{\theta}
J_{\theta} \rangle $, where the average is taken over the unit
cell, and can be shown to be $M = -(p A_0/4\pi) (1 - 1/\epsilon)
I_2(\omega R/c)/I_0(\omega R/c)$.

The effective permittivity and permeability have been calculated
for a range of wavenumbers $\vec{k}= k\vec{e}_x$ and the
corresponding frequencies $\omega(k)$.
For $kd \ll \pi$ it follows
from the analyticity of $\omega(\vec{k})$ that the frequency
depends only on $|\vec{k}|$ and not on its direction.
For $k_0 = 0.6/d$ and $\omega_0 = 0.6 \omega_p$ (or $n_{eff} =
-1$) we numerically computed that $\mu_{eff} = -2.35$ and
$\epsilon_{eff} = -0.427$. Therefore, at this frequency our SPC is
a DNM. Note that the hybridized monopole/octupole resonance
affects not only the magnetic permeability of the SPC, but also
the dielectric permittivity: the mean-field calculation using
Eq.~(\ref{eq:eps_eff1}) yields $\epsilon_{qs}(\omega_0) = -0.65$
that is significantly different from $\epsilon_{eff}$.
\begin{figure}
\includegraphics[height=5.2cm]{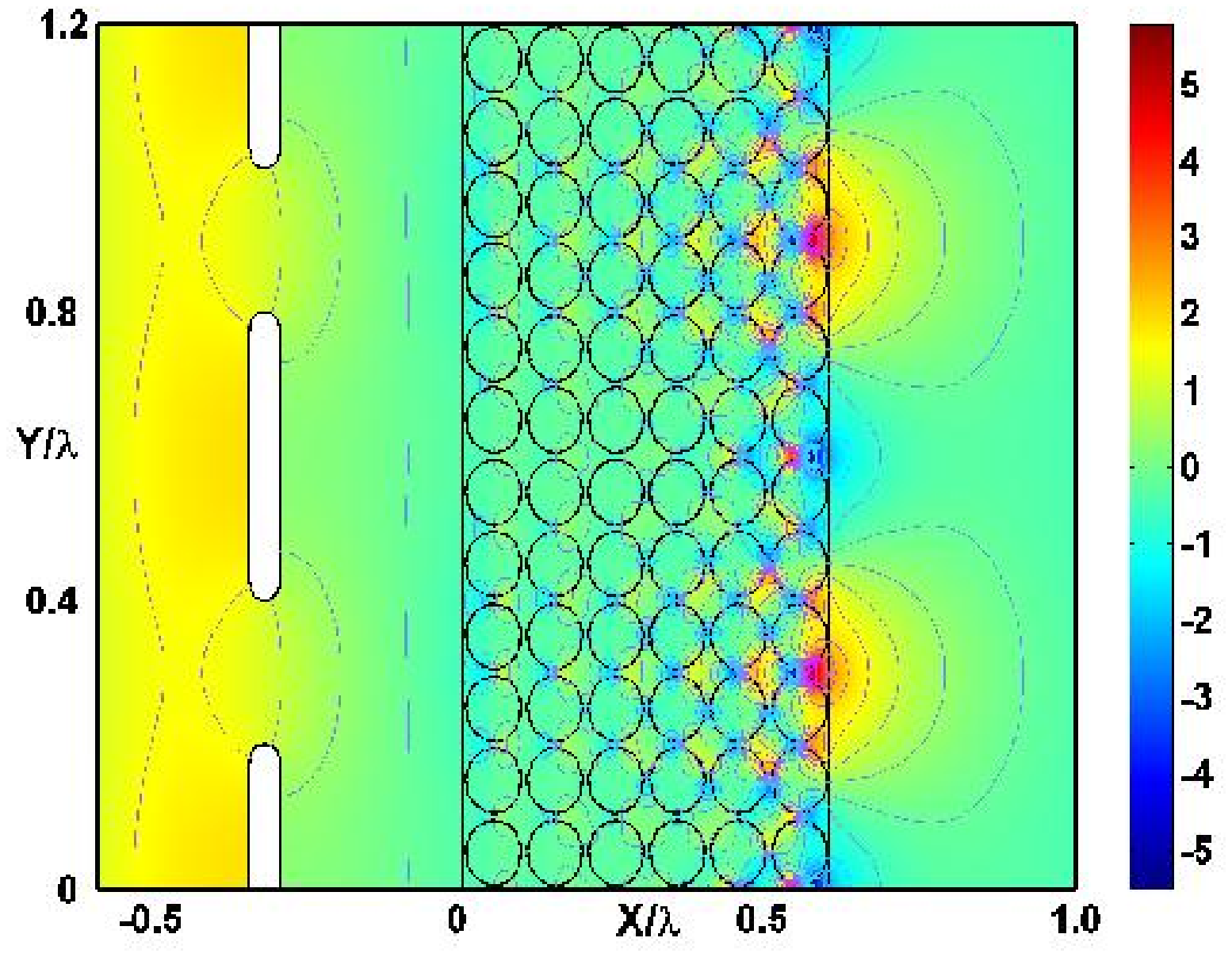}
\includegraphics[height=5.2cm]{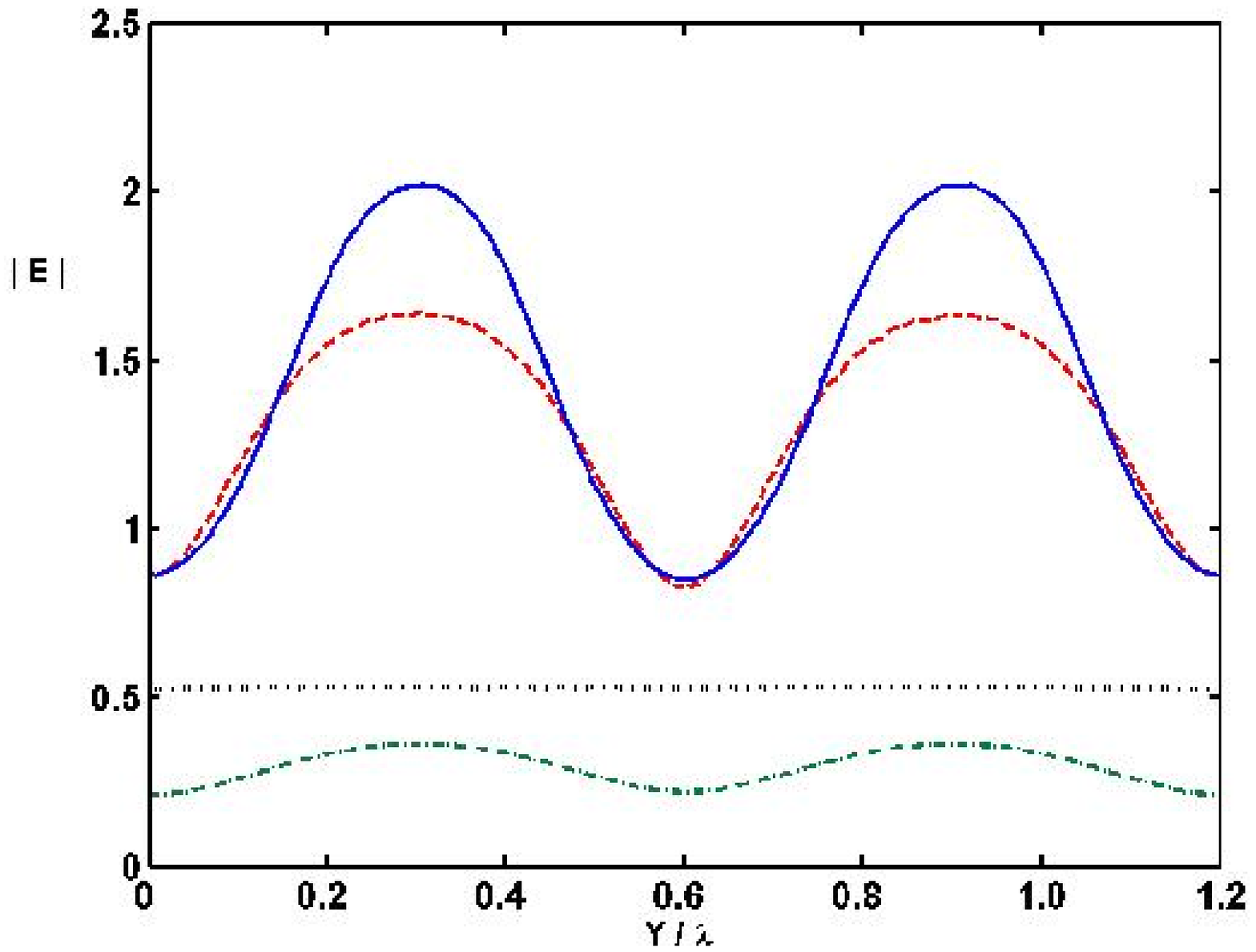}
\caption{\label{fig:super_lens} (a) Magnetic field distribution
behind an illuminated periodic slit array, with a six-period SPC,
parameters as in Fig.~\ref{fig:two_resonances}. (b) $|\vec{E}|$ in
the object plane (solid line); in the image plane for $\omega_0 =
0.6 \omega_p$, without damping (dashed line) and with damping
characteristic of silver (dot-dashed line); in the image plane for
$\omega = 0.606 \omega_p$ (dotted line). }
\end{figure}

DNM-based flat super-lenses capable of sub-wavelength imaging have
been proposed~\cite{pendrylens_prl00}. The condition for
super-lensing is that the DNM with the dielectric permittivity
$\epsilon < 0$ is embedded in a host medium with $\epsilon_h = -
\epsilon$. We have tested a six-period thick plasmonic SPC for the
super-lensing effect by embedding it inside the hypothetic host
with $\epsilon_h  = 0.55$. This particular choice of
$-\epsilon_{eff} < \epsilon_h < -\epsilon_{qs}$ was not optimized,
and is one among the several that showed super-lensing. To verify
the sub-wavelength resolution, we simulated the distribution of
the magnetic field $|\vec{E}|$ behind a screen with narrow slits
of width $\Delta_y = \lambda/5$ separated by a distance $2
\Delta_y$. As depicted in Fig.~\ref{fig:super_lens}, where only
two slits are shown, a planar wave with frequency $\omega = 0.6
\omega_p$ is incident on the screen from the left. A six-period
long plasmonic SPC of width $D = 0.6 \lambda$ is positioned
between $0 < x < D$. The distribution of $|\vec{E}|$ in the $x-y$
plane is shown in Fig.~\ref{fig:super_lens}(a). Also, in
Fig.~\ref{fig:super_lens}(b) $|\vec{E}|$ is plotted in two
cross-sections: the object plane right behind the screen (at $x =
- D/2 + \lambda/10$, solid line), and in the image plane (at $x =
3D/2 - \lambda/10$, dashed line). Object plane is slightly
displaced from the screen to avoid $\vec{E}$-field spikes at the
slit edges. The two sub-wavelength slits are clearly resolved.
Increasing the incident frequency by just one percent (outside of
the DNM band) results in the complete loss of resolution in the
image plane (dotted line).

While the DNM band for the plasmonic SPC is quite narrow,
$[\omega(k=0) - \omega(k=\pi/d)]/\omega(k=0) = 0.055$, it is still
broader than the collisional linewidth for some plasmonic
materials. For example, for silver $\epsilon = \epsilon_b -
\omega_p^2/\omega (\omega + i\gamma)$, where $\epsilon_b \approx
5$, $\omega_p = 9.1$ eV, and $\gamma = 0.02$
eV~\cite{christy_prb72}. Fig.~\ref{fig:super_lens}(b) (dash-dotted
line) confirms that, although finite damping $\gamma/\omega_p =
0.002$ reduces the field amplitude in the image plane, it does not
affect the image contrast. The band flatness in a plasmonic SPC
translates into very sharp excitation resonances, and large
enhancements of the incident field. For example, the very close
proximity of two flat propagation bands in
Fig.~\ref{fig:super_lens}(b) (second DM and the $A_1$ HMM) can be
exploited for maximizing the structure response at the incident
and Raman-shifted re-emitted frequencies, which is essential for
SERS.

This work is supported by the NSF's Nanoscale Exploratory Research
Contract No.~DMI-0304660.

\bibliography{photobib}

\end{document}